\documentclass[letterpaper,english,american,aps,prx,twocolumn,amsmath,amssymb,showpacs,superscriptaddress,longbibliography]{revtex4-2}
\usepackage{babel}
\usepackage{prettyref}
\usepackage{mathtools}
\usepackage{amsmath}
\usepackage{amssymb}
\usepackage{graphicx}
\usepackage[unicode=true,pdfusetitle,
 bookmarks=true,bookmarksnumbered=false,bookmarksopen=false,
 breaklinks=true,pdfborder={0 0 1},backref=false,colorlinks=false]
 {hyperref}
\hypersetup{
 colorlinks=true,linkcolor=black,citecolor=black,urlcolor=black,filecolor=black}
\usepackage{breakurl}
\usepackage{comment}
\makeatletter

\pdfpageheight\paperheight
\pdfpagewidth\paperwidth

\usepackage{tikz}
\usetikzlibrary{calc}

\newrefformat{cap}{\hyperref[#1]{Fig.~\ref{#1}}}
\newrefformat{fig}{\hyperref[#1]{Fig.~\ref{#1}}}
\newrefformat{tab}{\hyperref[#1]{Table ~\ref{#1}}}
\newrefformat{sec}{\hyperref[#1]{Sec.~\ref{#1}}}
\newrefformat{sub}{\hyperref[#1]{Sec.~\ref{#1}}}
\newrefformat{cha}{\hyperref[#1]{Chap.~\ref{#1}}}
\newrefformat{app}{\hyperref[#1]{App.~\ref{#1}}}

\makeatletter
\renewcommand{\fnum@figure}{FIG. \thefigure}
\makeatother

\usepackage{dsfont}

\makeatother

\begin{document}
\global\long\def\erf{\mathrm{erf}}%
\global\long\def\T{\mathrm{T}}%
\foreignlanguage{english}{}
\global\long\def\x{\mathbf{x}}%
\foreignlanguage{english}{}
\global\long\def\sgn{\mathrm{sign}}%
\foreignlanguage{english}{}
\global\long\def\Tr{\mathrm{Tr}}%
\title{Global hierarchy vs. local structure: spurious self-feedback in
scale-free networks}
\author{Claudia Merger}
\affiliation{Institut für Theoretische Festkörperphysik, RWTH Aachen University, 52056 Aachen
Germany}
\author{Timo Reinartz}
\affiliation{Institut für Theoretische Festkörperphysik, RWTH Aachen University, 52056 Aachen
Germany}
\author{Stefan Wessel}
\affiliation{Institut für Theoretische Festkörperphysik, RWTH  Aachen University, 52056 Aachen
Germany}
\author{Carsten Honerkamp}
\affiliation{Institut für Theoretische Festkörperphysik, RWTH Aachen University, 52056 Aachen
Germany}
\author{Andreas Schuppert}
\affiliation{Aachen Institute for Advanced Study in Computational
Engineering Science(AICES) Graduate School,
RWTH Aachen University, Germany and
Joint Research Center for Computational Biomedicine(JRC-Combine),
RWTH Aachen University, Germany}
\author{Moritz Helias}
\affiliation{Institute of Neuroscience and Medicine (INM-6) and Institute for Advanced
Simulation (IAS-6) and JARA BRAIN Institute I, Jülich Research Centre,
Jülich, Germany}
\affiliation{Institut für Theoretische Festkörperphysik, RWTH Aachen University, 52056 Aachen
Germany}
\date{\today}

\begin{abstract}
Networks with fat-tailed degree distributions are omnipresent across many scientific disciplines.
Such systems are characterized by so-called hubs, specific nodes with high numbers of connections to other nodes. By this property, they are expected to be key to 
the collective network behavior, e.g., in Ising models on such complex topologies.
This applies in particular to the transition into a globally ordered network state, which thereby proceeds in a hierarchical fashion, and with a non-trivial local structure. 
Standard 
mean-field theory of Ising models on scale-free networks underrates the presence of the hubs, while nevertheless
providing remarkably reliable estimates for the onset of global order. Here, we
expose that a spurious self-feedback effect, inherent to mean-field theory,
underlies this apparent paradox. More specifically, we demonstrate that higher order interaction effects precisely cancel the self-feedback on the hubs, and we expose
the importance of hubs for the distinct onset of local versus global order in the network.
Due to the generic nature of our arguments, 
we expect the mechanism that we uncover for the archetypal case of Ising networks of the Barabási-Albert type to be also relevant for 
other systems with a strongly hierarchical underlying network structure.
    
\end{abstract}
\maketitle
\section{Introduction}

Hierarchical networks  are found ubiquitously  across many areas
of physics,  as well as in social interaction clusters,  biological systems, and   medicine~\citep{kumar_web_2000,tanaka_scale-rich_2005,goldenberg_survey_2009,krogan_global_2006,bianconi_message-passing_2021,PhysRevResearch.3.013070}.
While the full universality of scale-free or power-law behavior in the underlying
connectivity structure is still under debate \citep{broido_scale-free_2019,holme_rare_2019}, many real-world networks clearly exhibit  strong heterogeneities.
Such networks are often also characterized by the presence of hubs, i.e. nodes of high degree, which strongly influence the network's properties.
Highly connected hubs are also expected to be important for the collective behavior in networks of node-based agents that interact through the network's topology.

In order to examine the impact that hubs have on both the global behavior as well as the local properties in such systems, we here examine a specific statistical mechanics model on a hierarchical network. 
For this purpose, we consider the nodes of the network to consist
of binary units that interact across a scale-free network structure.
In particular, we examine the 
Ising model on Barabási-Albert (BA) networks \citep{barabasi_emergence_1999}.
While in this model all interactions are set to the same strength, the number
of connections of a given node $i$, also called the node's degree $k_i$, varies strongly across the network,
defining thereby a hierarchy on the network topology. We investigate how this
hierarchy influences the emergence of order in especially
the presence of hubs whose degree may even scale with overall size of the network. 

Early studies of
Barabási-Albert-Ising  (BAI) networks have 
been performed in particular using Monte Carlo
simulations~\citep{herrero_ising_2004,aleksiejuk_ferromagnetic_2002}
and mean-field theory~\citep{bianconi_mean_2002}.
These studies demonstrated that for finite network sizes (in terms of the number of nodes $N$) a strong alignment of the Ising degrees of freedom emerges below a specific temperature, resembling the onset of ferromagnetic order in conventional lattice Ising models. In the conventional case, the ferromagnetically ordered phase emerges below a finite transition temperature out of the paramagnetic high-temperature phase in the thermodynamic (large-$N$) limit. 
For the BAI model, however,  Monte Carlo simulations indicated that the effective transition temperature 
$T_{T}$ instead grows logarithmically with the network size $T_{T}\propto\log(N)$.
In view of this numerical finding, Bianconi \citep{bianconi_mean_2002} then presented a mean-field
theory of BAI models, describing the network state in terms of a single global order parameter \citep{bianconi_mean_2002}. This approach indeed reproduces the observed logarithmic scaling of the effective transition temperature.
A linearization of
the mean-field self-consistency equations near the transition temperature furthermore shows that the average magnetization
$m_{i}$ of a given node $i$ increases proportional to its degree
\begin{align}
m_{i} & \propto k_{i}.\label{eq:mi_prop_ki}
\end{align}
This implies a rather simple local structure, in which the magnetization is determined solely by the degree. Other studies that analyzed the emergence of order in the Ising model on scale-free networks employed recursion methods for tree-like networks \citep{dorogovtsev_ising_2002} or the replica trick \citep{leone_ferromagnetic_2002}. All previous studies reproduced the logarithmic scaling of the transition temperature with the network size, while 
they characterized the magnetic order in terms of the degree of the nodes, as in the mean-field approach. 

Here, we demonstrate that the mean-field theory of Ref.~\citep{bianconi_mean_2002}, even though it predicts the scaling of the global ordering transition, does not accurately describe the local structure in the collective network behavior. In particular, as one of our main findings, we report results from refined Monte Carlo simulations, which exhibit that the hubs enforce a noticeable stronger local magnetization on their direct neighbors as compared to a typical node. The value of a node's magnetization $m_i$ is thus far from being determined by its degree $k_i$ alone. This behavior stems from the presence of a strong local alignment field $h_u$, where $u$ denotes one of the hubs, that acts on the nearest neighbors of each hub. 
In fact, our simulations expose the presence of an additional shell-like structure in the magnetization profile, which is not captured in the above-mentioned mean-field description. In view of its ignorance with respect to the actual local structure, while still accurately capturing the global onset of order in terms of a single global order parameter, we denote the mean-field theory of Ref.~\citep{bianconi_mean_2002} as a global mean-field theory in the following. 

Alternatively, one can perform a mean-field calculation on the level of individual Ising variables, including the actual local structure. However, for this analysis we find that the local mean-field theory severely overestimates the role of hubs. In particular, a hub's local alignment field $h_u$ induces a strong magnetization on its nearest neighbor nodes, which, in turn raises the magnetization of the hub itself. The hub thus effectively feels its own field. Due to this self-feedback, the local mean-field approach predicts a much higher transition temperature than actually observed, scaling proportionally to $N^{\frac{1}{4}}\gg\log(N)$. Moreover, the magnetization of these  ordered states are strongly localized around the hubs, in contradiction to the simple proportionality relation \eqref{eq:mi_prop_ki}. The ordered states are therefore also highly sensitive to the presence of individual hubs in the network. This makes them fragile in terms of the  stability to small perturbations in the network topology, such as the random removal of nodes. This sensitivity also appears to be in conflict with the collective nature of the onset of global order.
The contrast between the global and the local mean-field theory ultimately exposes an inherent inconsistency of the degree-resolved approach to heterogeneous systems.

As we report below, this problem can be overcome upon expanding beyond mean-field theory and analyzing the BAI using a self-consistent Thouless-Anderson-Palmer (TAP)  approach~\citep{fischer_spin_1991,vasiliev_legendre_1974,thouless_solution_1977}, which takes second order interaction effects into account. While the TAP approach has been applied to various systems in the past, including, e.g., spin glasses, we are not aware of of any previous application of the TAP approach to scale-free hierarchical networks. 

Within the TAP framework, we identify an explicit cancellation of the self-feedback on the hubs on the microscopic level due to the fluctuations that are not captured by the mean-field approach. Moreover, this cancellation is found to be independent of the specific network structure, and is indeed reminiscent of the well-known cavity argument \citep{mezard_spin_1986}: 
The local alignment field $h_i$ at a given node of the network needs to be calculated in the absence of this node. In consequence, the local fields caused by the hubs emerge only when the system orders globally, and  are much weaker than predicted by the local mean-field theory.
Upon taking this effect into account, the TAP approach yields the correct scaling of the transition temperature $T_{T}\sim\log(N)$, while at the same time it maintains the full information on the network connectivity and thus explains the hierarchical  structure of the local magnetization. The theoretical framework that we put forward also allows us to obtain accurate predictions of various thermodynamic quantities that compare well to numerical data obtained by Monte Carlo simulations.

Monte Carlo simulations of heterogeneous networks based on local update schemes are impeded by the freezing of the hub's magnetic states. We find that cluster update schemes do not efficiently alleviate this problem due to the high connectivity of the hubs, and thus still  suffer from large autocorrelation times. 
In our simulations, we thus used an improved version of parallel-tempering Monte Carlo instead \citep{katzgraber_feedback-optimized_2006} that allows us to perform controlled simulation by adequately sampling over the full configuration space. In particular, this approach renders an accurate picture of the behavior of hubs and their surrounding nodes throughout the entire relevant temperature range. 

The remainder of this paper is organized as follows: In~\prettyref{sec:Hierarchical-Model}
we introduce the BAI model. \prettyref{sec:Monte-Carlo} defines the observables studied here and describes how they can be measured from finite-size Monte Carlo simulations.
Then, in \prettyref{sec:The-conundrum-Mean-field-theory} we expose the inherent inconsistency of the mean-field theory, and
resolve it in \prettyref{sec:Self-feedback-and-TAP}, by showing that the self-feedback effect is canceled by fluctuations. We demonstrate how an effective coupling to a global ordering field approximates the network state well, which can be described in terms of degree-resolved magnetizations. Further local information is then included to refine this approximation.
We compare observations from finite size Monte Carlo and TAP theory in \prettyref{sec:Finite-size-effects}.
Finally, \prettyref{sec:Discussion} provides a summary and further discussion of our findings, and also presents an outlook on how we expect them to generalize to other models.

\section{Hierarchical BAI Model\label{sec:Hierarchical-Model}}
The statistical mechanics  model that we consider in the following is defined on a BA network~\citep{barabasi_emergence_1999}. Quite generally, BA networks are specific hierarchical undirected networks,  which can be generated ("grown") using an iterative, stochastic algorithm (see \prettyref{app:Construction-algorithm} for details), starting from $m_0$ initial nodes, one consecutively adds new nodes, where each new node is connected to $m_0$ other nodes of the network. During the growth process of a BA network, nodes that already have a large degree are most likely  to form a new connection to the newly added node -- a principle known as preferential attachment. The final network
of size $N$ is then specified by its $N\times N$ adjacency matrix $\mathbf{A}$, where for nodes $i,j\in\{1,...N\}$ 
\begin{align}
A_{ij}=A_{ji}= & \begin{cases}
1 & \text{if \ensuremath{i\neq j} and }i,j\text{ are connected}\\
0 & \text{ else}.
\end{cases}\label{eq:Adjacency_entries}
\end{align}

It was shown in Ref.~\citep{barabasi_emergence_1999} that in the large-$N$ limit
these networks feature a scale-free degree statistics, i.e., the probability that a randomly chosen node has a specific degree $k$ is given by
\begin{align*}
    p(k)= & \frac{2m_{0}^{2}}{k^{3}}.
\end{align*}

Furthermore, in the same limit, the BA network exhibits linear degree-degree correlations \citep{bianconi_mean_2002}, where 
the probability that two nodes $i$ and $j$ are connected is given as 
\begin{align}
p_{c}(k_{i},k_{j}) = & \frac{k_{i}k_{j}}{2m_{0}N}\label{eq:p_c_as_k}
\end{align}
in terms of their degrees $k_i$ and $k_j$, respectively. 
By construction (see \prettyref{app:Construction-algorithm}) the average degree across all nodes of a realization of a BA network is
\begin{align}
\langle k\rangle = & 2\,m_{0}.\label{eq:mean_degree}
\end{align}
The largest degree, averaged over multiple realizations of finite BA networks of the same size, is given by
\begin{equation}
k_{\mathrm{max}}=m_{0}\sqrt{N} .
\end{equation}

As a simple statistical physics model of how agents interact on strongly hierarchical networks, we now consider an Ising model on the BA network: to each node $i\in{1,\dots,N}$ a local binary degree of freedom (spin) $x_{i}\in\{-1,1\}$ is assigned, and the probability for a given configuration $x=(x_1,...,x_N)$ is determined by the  Boltzmann-factor $p(\mathbf{x})\propto\exp\left(-\beta H(\mathbf{x})\right)$ in terms of the configuration's energy $H(\mathbf{x}) = -\frac{J}{2}\,\mathbf{x} \cdot A\,\mathbf{x}$. The ferromagnetic coupling $J$ in $H$ favors the alignment of connected spins into
either direction, and $\beta=T^{-1}$ is the inverse temperature ($k_B=1$),
measured in relation to the elementary energy scale $J$, which we therefore fix to $J=1$ in the following: 

\begin{align}
H(\mathbf{x},\mathbf{h}) & =-\frac{1}{2}\,\mathbf{x} \cdot \mathbf{A} \,\mathbf{x}-\mathbf{h}\cdot \mathbf{x}.\label{eq:H(x)}
\end{align}

In addition, an  external magnetic field $\mathbf{h}$ is included in the Hamiltonian $H$, which will be set to zero for our further analysis. However, the inclusion of such a term will turn out convenient for the formulation of the TAP theory (where $\mathbf{h}$ will be considered an infinitesimal source term).

The physical properties of the BAI model are finally defined in terms of statistical averages over all spin configurations, $\langle O\rangle = \sum_{\mathbf{x}} O(\mathbf{x}) \, p(\mathbf{x})$, of appropriate observables $O$, specified further below. Since performing the summation over all $2^{N}$  spin configurations is unfeasible for large $N$, here we use parallel-tempering Monte Carlo simulations to calculate these statistical averages (see. \prettyref{app:Monte-Carlo} for details regarding the employed simulation scheme). In order to compare finite-size Monte-Carlo simulations to results obtained from the self-consistent mean-field and TAP approaches, we consider appropriately defined Monte Carlo observables, as discussed in the following section. 

\section{Monte Carlo observables on finite networks}
\label{sec:Monte-Carlo}

When studying magnetic systems that undergo order-disorder phase transitions, one is typically  interested in the thermodynamic limit, i.e.,  one  considers the limit of infinite system sizes, as only in this limit the spontaneous symmetry breaking associated with the onset of order can emerge (see, e.g, the discussion in \cite{goldenfeld_lectures_1992}, Sec. 2).
In finite systems, one can nevertheless detect the emergence of  magnetic order in terms of an effective transition temperature that approaches the true ordering transition temperature in the thermodynamic limit. Such effective transition temperatures can be defined based on various thermodynamic quantities, such as the peak position of the specific heat or the magnetic susceptibility. A detailed discussion of this approach for the Ising model on a regular lattice geometry can be found, e.g., in Ref.~\citep{landau_guide_2005}.
 
For the BAI, the effective transition temperature scales logarithmically with the network size~\cite{herrero_ising_2004,aleksiejuk_ferromagnetic_2002}. In the thermodynamic limit the system thus resides in the ordered state at all finite temperatures. 
One thus faces an interesting dichotomy: The concept of a phase transition only acquires meaning in the thermodynamic limit, but it is precisely the infinite network that does not exhibit this phenomenon, due to the diverging effective transition temperature. 
On the other hand, despite this dichotomy, one can take a more pragmatic view and try to describe a large but finite system as well as possible. This should be of relevance to most applications of the model.

More generally, one must proceed with care when comparing results from mean-field theory and the TAP approach to  Monte Carlo simulations of finite systems. 
In particular, from ergodic Monte Carlo simulations, one cannot obtain a finite value of the mean magnetization that characterizes the magnetic state of the mean-field and TAP approach. Instead, $\langle x_{i}\rangle=0$ vanishes exactly for a sufficiently long run, due to the $Z_{2}$-symmetry of the Ising model Hamiltonian, which for a finite system also implies the exact vanishing of the zero-field  magnetization
\begin{equation}
M_{0}
=\frac{1}{N}\bigg\langle  \sum_{i} x_{i} \bigg\rangle\,,
\label{eq:Disordered-Magnetization}
\end{equation}
that enters the explicit form of  the zero-field susceptibility
\begin{equation}
    \chi_{0} = \frac{dM(h)}{dh}\bigg|_{h=0} =\beta \bigg( \frac{1}{N} \sum_{i,j}  \langle x_i x_j \rangle - N\, ( M_{0} )^2 \bigg) \,.
    \label{eq:Susceptibility}
\end{equation}
For the Ising model on regular lattices with a finite  transition temperature $T_T$ in the thermodynamic limit, this quantity converges to the zero-field susceptibility in the thermodynamic limit only for temperatures $T>T_T$~\citep{landau_guide_2005}. 

For temperatures below the  transition temperature, the mean value of the absolute magnetization, which we denote by 
\begin{equation}
M_\mathrm{MC}
=\frac{1}{N}\bigg\langle \bigg| \sum_{i} x_{i} \bigg| \bigg\rangle\,,
\label{eq:Absolute-Magnetization}
\end{equation}
can instead be used to probe a unimodal symmetry broken state and thus also allows us to compare to the magnetic solutions of mean-field and TAP theory.  In terms of $M_\mathrm{MC}$, we also consider the estimator
\begin{equation}
    \chi_\mathrm{MC} =\beta \bigg( \frac{1}{N} \sum_{i,j}  \langle x_i x_j \rangle - N \, (M_\mathrm{MC})^2 \bigg) \,,
    \label{eq:Susceptibility_MC}
\end{equation}
which is appropriate to calculate the susceptibility within the symmetry broken regime~\citep{landau_guide_2005}.
One can also express $M_\mathrm{MC}$ in terms of the orientation of the individual spins with respect to the overall magnetization, since
\begin{equation}
M_\mathrm{MC}=\frac{1}{N}\sum_i m_{i,\mathrm{MC}},
\end{equation}
where $m_{i,\mathrm{MC}}$ quantifies the alignment of individual spins with the overall magnetization,
\begin{equation}
    m_{i,\mathrm{MC}} = \bigg\langle x_i \, \sgn \big( \sum_j x_j \big) \bigg\rangle\,.
    \label{eq:MC_mag_h}
\end{equation}

In addition to these magnetic properties, we also consider the energy
\begin{equation}
    E = \frac{1}{N}\langle H\rangle,
    \label{eq:Energy}
\end{equation}
and the specific heat
\begin{equation}
    C =\frac{d E}{d T} = \frac{\beta^2}{N} \big(\langle H^2 \rangle- \langle H \rangle^2\big) \,
    \label{eq:SpecificHeat}
\end{equation}
of the BAI system.
Based on the  quantities introduced above, the Monte Carlo simulations allow us to benchmark the accuracy of the TAP approach, as well as to expose the severe limitations of earlier mean-field theories. In the next section, we start by revisiting the mean-field approach.

\section{The conundrum: Mean-field theory\label{sec:The-conundrum-Mean-field-theory}}

Within the mean-field approximation~\citep{bianconi_mean_2002}, the exact equation
\begin{equation}
\langle x_{i}\rangle=\bigg\langle \tanh\bigg(\beta\,\sum_{j}A_{ij} x _{j}\bigg)\bigg\rangle\label{eq:exact}
\end{equation}
for the average value of the magnetization $m_{i}=\langle x_{i}\rangle$ at node  $i=1,...,N$ is replaced by the self-consistency equation
\begin{equation}
m_{i}=\tanh\bigg(\beta\,\sum_{j}A_{ij}\, m_{j}\bigg).
\label{eq:MF}
\end{equation}
The trivial solution $m_{i}=0$, $i=1,...,N$ exists for all values of $\beta$. Here, we are interested in non-trivial solutions to \prettyref{eq:MF}, for which $\sum_{i}|m_{i}|>0$. Such solutions feature a non-zero value of (at least one of) the variables $m_i$, and always come in pairs, since \prettyref{eq:MF} is symmetric under the simultaneous inversion of the sign of all $m_i$. These non-trivial solutions are found for values of $\beta$ larger than a particular value $\beta_T$, which defines the mean-field estimate for the effective transition temperature $T_T=1/\beta_T$ for the onset of the magnetic order.

In the following, we will analyze the mean-field equations~\prettyref{eq:MF} in two different ways: First, within the local mean-field theory approach, we continue to treat all spins on the microscopic level in terms of the individual average magnetization $m_{i}$. Second, within the global mean-field theory approach, we consider only a single global order parameter, denoted $S$, to be defined further below. We will derive for $S$ a self-consistency equation on the global network level. Its solution  with a finite value of $S$ thus indicates the presence of magnetic order in the network. 

We will show below that (i) the two approaches yield inconsistent results, (ii) only the
latter is in agreement the with effective transition temperature scaling obtained from Monte Carlo simulations, and (iii) it nevertheless fails to account for the local shell-like structure present in the magnetic state of the BAI model.

\paragraph{Local mean-field theory.}
On a finite network, non-trivial solutions of the local mean-field self-consistency equations \prettyref{eq:MF} emerge in a continuous manner, such that $\sum_i|m_{i}|\rightarrow0$ continuously upon tuning $\beta\rightarrow\beta_{T}$ from above. We can thus linearize \prettyref{eq:MF} near $\beta_T$ to obtain 
\begin{equation}
m_{i}=\beta_{\mathrm{MF}}^{\text{local}}\,\sum_{j}A_{ij}m_{j},\label{eq:MF_linear}
\end{equation}
where  $\beta_{\mathrm{MF}}^{\text{local}}$ denotes the value of $\beta_T$ within the local mean-field theory approach. Finding the transition temperature therefore reduces to the calculation of the largest eigenvalue $\lambda_{\mathbf{A},\mathrm{max}}$ of the adjacency matrix
\begin{equation}
\beta_{\mathrm{MF}}^{\text{local}}=\lambda_{\mathbf{A},\mathrm{max}}^{-1}.\label{eq:MF_True_T_T}
\end{equation}
Here, only the largest eigenvalue of $\mathbf{A}$ yields a valid solution, since the linearized equation \prettyref{eq:MF_linear} applies only at the transition temperature, below which a solution with finite $\sum_i |m_{i}|>0$ emerges. Since $\mathbf{A}$ is real and symmetric, its largest eigenvalue is equal to the matrix norm $||\mathbf{A}||$. In Ref.~\citep{goh_spectra_2001}, the scaling of the largest eigenvalue was found to be $\lambda_{\mathbf{A},\mathrm{max}}\propto \sqrt{m_{0}}\,N^{{1}/{4}}$, for which  the corresponding eigenvector is strongly localized at the node of highest degree. It is easy to see that indeed the matrix norm $||\mathbf{A}||$ scales as $N^{{1}/{4}}$: To this end, consider a vector $\mathbf{v}$ of unit length that takes on the  value $v_{u}={1}/{\sqrt{2}}$ for a  specific node $u$, while  $v_{i}={A_{iu}}/{\sqrt{2k_{u}}}$ for all other nodes, $i\neq u$. This vector is therefore non-zero only on node $u$ and its nearest neighbors. Then
\begin{align*}
\lambda_{\mathbf{A},\mathrm{max}}=||\mathbf{A}||\geq & ||\mathbf{A\,v}||\\
\geq & \sqrt{\left(\sum_{i}A_{ui}\frac{A_{iu}}{\sqrt{2k_{u}}}\right)^2}
=  \sqrt{\frac{k_{u}}{2}},
\end{align*}
where we used that $\sum_{i}A_{ui}^{2}=k_{u}$ equals the degree of node $u$. Now choosing $u$ to be the node with the largest degree $k_{\mathrm{max}}\propto\sqrt{N}$ \citep{barabasi_emergence_1999}, we obtain $\lambda_{\mathbf{A},\mathrm{max}}\propto N^{{1}/{4}}$. In conclusion, within the local mean-field theory, a non-vanishing solution of \prettyref{eq:MF_linear} exists below a temperature $T_{\mathrm{MF}}^{\text{local}}$, that scales as 
\begin{align}
T_{\mathrm{MF}}^{\text{local}} & \propto N^{{1}/{4}}\gg\log(N)\label{eq:T_transition_lambda}
\end{align}
for large $N$. This result differs significantly from the transition temperature to the ferromagnetic state at $T_T\propto \log(N)$, predicted by the global mean-field theory with a global order parameter \citep{bianconi_mean_2002}, which will be revisited next, and it also contradicts the results from  Monte Carlo simulations \citep{herrero_ising_2004,aleksiejuk_ferromagnetic_2002}.

\paragraph*{Global mean-field theory.}

To demonstrate that the global mean-field theory approach with a single global order parameter \citep{bianconi_mean_2002} yields a different scaling than \prettyref{eq:T_transition_lambda}, we briefly review its derivation here: 
The probability of two nodes to be connected is expressed by their respective degrees \prettyref{eq:p_c_as_k}. 
One then proceeds by replacing the adjacency matrix $A_{ij}$ in \prettyref{eq:MF_linear} by the corresponding probability $p_{c}(k_{i},k_{j})$. Within this approximation, nodes of the same degree hence obey the same mean-field equations, so that the value of the local magnetization depends only on the degree of the node, $m_{i}=m(k_{i})$. One can therefore replace a summation $\sum_{i=1}^{N}$ over all nodes by a summation over the degrees $\sum_{k}N\,p(k)$. Applied to \prettyref{eq:MF_linear} one finds that 
\begin{equation}
m(k_{i})=k_{i}\beta\underbrace{\frac{1}{2m_{0}}\sum_{k}p(k)\,k\,m(k)}_{=S}\propto k_i,\label{eq:MF_Linear_Degree_Scaling}
\end{equation}
where we introduced the global order parameter
\begin{equation}
S=\frac{1}{2m_{0}}\sum_{k}p(k)\,k\,m(k).
\end{equation}
 Using  \prettyref{eq:mean_degree} this global order parameter obeys the equation
\[
S=\frac{\langle k^{2}\rangle}{\langle k\rangle}\beta\,S,
\]
from which the transition temperature of the global mean-field theory approach follows as
\begin{equation}
T_{\mathrm{MF}}^{\text{global}}=\frac{\langle k^{2}\rangle}{\langle k\rangle}\approx\frac{m_{0}}{2}\,\log(N)\, . \label{eq:MF_Log_Temperature}
\end{equation}

Comparing \eqref{eq:T_transition_lambda} and \eqref{eq:MF_Log_Temperature} for large networks $N\gg1$,
we obtain
\begin{align*}
T_{\mathrm{MF}}^{\text{local}}
\,{\gg}\,T_{\mathrm{MF}}^{\text{global}}.
\end{align*}
This shows an inherent inconsistency of the mean-field approach: At the lower temperature $T_{\mathrm{MF}}^{\text{global}}\ll T_{\mathrm{MF}}^{\text{local}}$ the local mean-field theory predicts that the system resides already well within the regime for which a non-trivial solution exists. Therefore, the linearization \eqref{eq:MF_linear}, i.e., the starting point to derive \eqref{eq:MF_Log_Temperature}, has been employed outside its regime of validity.

To illustrate the previous point, we solve \prettyref{eq:MF} numerically for a fixed random realization of $\mathbf{A}$ by a fixpoint relaxation; a
positive solution (i.e., $m_{i}\protect\geq0$ for all $i$) is selected upon starting the relaxation from the fully polarized state (with $m_{i}=1$ for all $i$). Throughout this work, we analyze the same single realization of a BA network per system size. We find that this is sufficient because the characteristics of these networks vary only weakly across realizations for large system sizes (see \prettyref{app:Self-averaging-of-BA-networks}). We find these solutions to be typically strongly localized at the nodes of highest degree. We compare several characteristic observables obtained from the local mean-field theory with the results from Monte Carlo simulations.  We find that the average magnetization $M$
and energy $E$,
while showing a similar overall behavior at low temperatures,  exhibit different high-temperature behavior in \prettyref{fig:M,E}. In particular, \prettyref{fig:M,E}(b) demonstrates that non-zero solutions of \eqref{eq:MF} indeed emerge at higher temperatures than \prettyref{eq:MF_Log_Temperature}.
The local mean-field theory thus overestimates the tendency of the spins to order and it predicts the transition to a disordered state at a higher temperatures than the value obtained from the global mean-field theory, which matches better to the Monte Carlo data.

\begin{figure}
\begin{centering}
\includegraphics{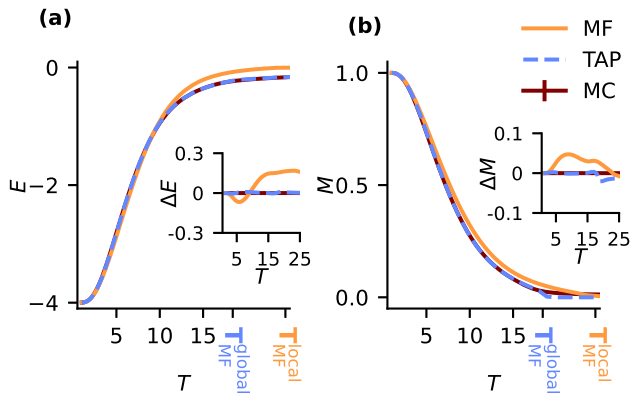}
\par\end{centering}

\caption{Energy $E$ (a) and magnetization $M$ (b) as functions of temperature for a BAI network with   $N=10^{4}$ and $m_{0}=4$. Results from local mean-field theory (MF) and  TAP are compared to Monte Carlo simulations (MC). 
All data are obtained for the same adjacency matrix $\mathbf{A}$.
The insets in both panels show the differences of MF and TAP results to MC.
\label{fig:M,E}}
\end{figure}
\begin{figure}
\begin{centering}
\includegraphics{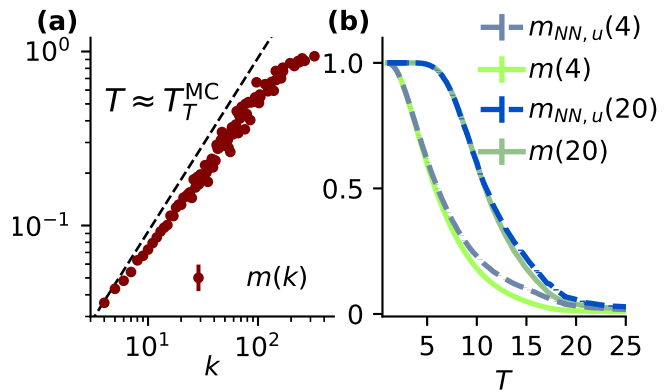}

\par\end{centering}
\caption{ (a) Degree-wise magnetization $m(k)$ as a function of the degree $k$ obtained
by Monte Carlo simulation for $N=10^{4}$ and $m_{0}=4$ at the effective transition
temperature $T_T^\mathrm{MC}=16.12 \pm 0.05$, as extracted from the position of the maximum in $\chi_\mathrm{MC}$. 
The dashed line is parallel to the identity, to show the approximate
linear relationship. (b) Local structure. Green curves: $m(k)$,
for $k\in\{4,20\}$ as a function of temperature. Blue dashed curves: Degree-wise
averages over the nearest neighbors of the
hub $u$ \prettyref{eq:DegreewiseMagofhubneighbours}.  \label{fig:Strong-local-differences-in-magnetization}}
\end{figure}

The origin of this discrepancy appears to be the use of $p_{c}(k_{i},k_{j})$, which eliminates the local structure of the network contained in $\mathbf{A}$. One could argue that valid solutions to \prettyref{eq:MF} can still be found if one excludes strongly localized solutions, which is the effect of replacing $\mathbf{A}$ by $p_{c}$ in the global mean-field theory. Indeed, we observe that below the saturation regime (i.e. for $m_{i,\mathrm{MC}}\ll 1$)  the average magnetization $m(k)$ of a node with a given degree $k$ 
\begin{equation}
    m(k)=\frac{1}{\sum_{i}\delta_{k_{i},k}}\sum_{i} \delta_{k_{i},k} \, m_i\,,
    \label{eq:DegreewiseMag}
\end{equation}
 from the Monte Carlo simulations, fulfills $m(k) \propto k$ cf.  \prettyref{fig:Strong-local-differences-in-magnetization}(a).
Therefore, there is a strong degree-dependent hierarchical structure in the resulting magnetization profile as predicted by global mean-field theory. 
On the other hand,  we find that the local magnetization at nodes of low degree is not characterized by the degree alone: In  \prettyref{fig:Strong-local-differences-in-magnetization}(b) we compare the average degree-wise magnetization \eqref{eq:DegreewiseMag}
of all nodes which have a given degree $k$ to the average magnetization of the nearest neighbors of a hub $u$, given by
\begin{equation}
    m_{\mathrm{NN}u}(k)=\frac{1}{\sum_{i}A_{iu}\delta_{k_{i},k}}\sum_{i}A_{iu}\, \delta_{k_{i},k}\, m_i.
    \label{eq:DegreewiseMagofhubneighbours}
\end{equation}
We find from this comparison that the hubs enforce a stronger ordering on their nearest neighbor nodes than predicted by the degree alone, $m_{\mathrm{NN}u}(k) > m(k)$. Therefore, the local structure cannot be entirely eliminated when describing the ordered network state, in contrast to the assumption entering the derivation of the global mean-field theory.

A consistent mean-field theory can therefore not predict \prettyref{eq:MF_Log_Temperature} without further assumptions made in the derivation, leading to the ignorance of the local network structure. We will resolve this contradiction in the following by showing that the transition temperature \prettyref{eq:MF_Log_Temperature} emerges naturally from the microscopic TAP equations, which simultaneously  also predict  the  local  structure  of  magnetization.

\section{Self-feedback and TAP equations} \label{sec:Self-feedback-and-TAP}

The TAP approach~\citep{fischer_spin_1991,vasiliev_legendre_1974,thouless_solution_1977} 
improves upon the mean-field theory \eqref{eq:MF} in the sense that it adds second order interaction effects. For this purpose, 
we start from  the free energy $F$, obtained from the sum over all spin configurations as
\begin{equation}
    e^{-\beta F(\mathbf{h}, \beta)} = \sum_\mathbf{x} e^{-\beta H(\mathbf{x};\mathbf{h})} \,.
    \label{eq: PartitionFunction}
\end{equation}
One can then compute the average local magnetizations $m_i$ from derivatives of the free energy with respect to $h_i$, where one sets $\mathbf{h}=0$ in the end. However, it is more convenient to define the Legendre-Fenchel transform $G(\mathbf{m},\beta)$ of the free energy
\begin{equation}
G(\mathbf{m},\beta)=\sup_{\mathbf{h}}F(\mathbf{h},\beta) + \mathbf{h}\cdot \mathbf{m}\,,\label{eq:G}
\end{equation}
thereby obtaining a thermodynamic potential which is a function of the mean local magnetizations $m_i$. It fulfills the equation of state
\begin{equation}
    \frac{d G}{d m_i} = h_i = 0\, .
    \label{eq:EquationOfState}
\end{equation}
In \prettyref{app:Plefka-expansion}, we demonstrate how to expand $G$ as a function of the nearest-neighbor couplings, thus in $\beta$. To second order, one obtains
\begin{align}
 \beta G(\mathbf{m},\beta) = & \frac{1}{2}\sum_{i} (1+m_{i})\ln\frac{1+m_{i}}{2}+(1-m_{i})\ln\frac{1-m_{i}}{2} \nonumber \\
 & -\frac{\beta}{2}\sum_{i\neq j}A_{ij}m_{i}m_{j} \nonumber \\
 & -\frac{\beta^2}{4}\sum_{i\neq j}A_{ij}(1-m_{i}^{2})(1-m_{j}^{2}) + \mathcal{O}\left(\beta^{3}\right)
 \,,
 \label{eq:G_expanded}
\end{align}
where (i) the term in the first line is the Shannon entropy of a set of independent binary variables, (ii)  the sum in the  second line, proportional to  $\beta$, takes the form of the inner energy in mean-field approximation, and  (iii) the term in the third line, proportional to  $\beta^2$, is known as the TAP or Onsager correction term \cite{thouless_solution_1977}.

The equation of state \prettyref{eq:EquationOfState} then takes the form of the TAP self-consistency equations
\begin{align}
m_{i} & =\tanh\bigg[\underbrace{\beta\sum_{j}A_{ij}m_{j}}_{\text{mean-field}}-\underbrace{\beta^{2}m_{i}\sum_{j}A_{ij}^{2}\left(1-m_{j}^{2}\right)}_{\text{TAP}}\bigg] \, , \label{eq:TAP}
\end{align}
containing terms of linear and quadratic order in $\beta$.
For a given adjacency matrix $\mathbf{A}$, we solve this equation numerically and compare the resulting averages to Monte Carlo simulations. We find that the average magnetization $M$ and inner energy (obtained using $E = N^{-1} \partial(\beta G )/\partial\beta$)  agree rather well with the numerical values (\prettyref{fig:M,E}). In particular, the deviations from Monte Carlo are overall smaller for TAP than for the local  mean-field theory.

To identify the transition temperature based on  the TAP approach, we examine the linearization of the  TAP self-consistence equation \prettyref{eq:TAP}, obtaining 
\begin{equation}
m_{i}=\beta\sum_{j}\left(A_{ij} - \beta\delta_{ij}k_{i} \right)\; m_j \,,\label{eq:TAP_linear}
\end{equation}
where $k_{i}=\sum_j A_{ij}$ is the degree of vertex $i$. The transition temperature is thus obtained in terms of the leading eigenvalue 
$\lambda_{\mathbf{{B}(\beta),\mathrm{max}}}$
of the  $\beta$-dependent matrix $\mathbf{B}(\beta)$, with
\begin{equation}
    B_{ij}(\beta)=A_{ij}-\beta\delta_{ij}k_{i}\,, \label{eq:def_B}
\end{equation}
from the condition that
\begin{equation}
\beta_{\mathrm{TAP}} = \lambda^{-1}_{\mathbf{{B}(\beta_{\mathrm{TAP}}),\mathrm{max}}}\,. \label{eq:beta_TAP}
\end{equation}
\begin{figure}
\begin{centering}
\includegraphics{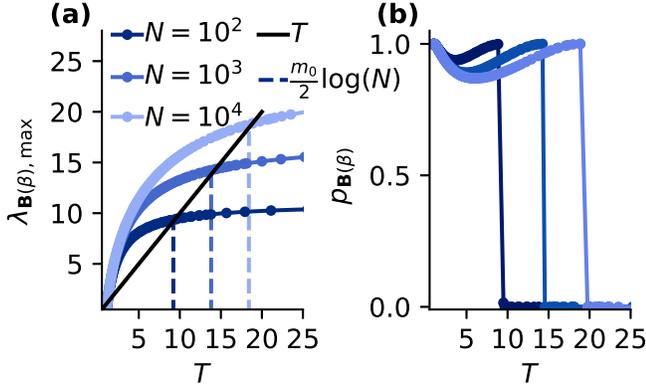}
\par\end{centering}
\caption{ (a) Leading eigenvalue $\lambda_{\mathbf{B}(\beta),\mathrm{max}}$ of $\mathbf{B}(\beta)$ \prettyref{eq:def_B} 
as a function of temperature $T$ for different system sizes $N$ and $m_{0}=4$.
The black solid line shows the identity and the vertical dashed lines are predictions for the point where  $\lambda_{\mathbf{B}(\beta),\mathrm{max}}=T$ from global TAP theory \eqref{eq:beta_TAP}. (b)
Projection $p_{\mathbf{B}(\beta)}$ \eqref{eq:ProjB} of the magnetization obtained from solving the TAP equations onto the leading eigenvector of $\mathbf{B}(\beta)$ for different system sizes $N$.\label{fig:Projection-B} \label{fig:Leading-eigenvalue-of-B}}
\end{figure}
In \prettyref{fig:Leading-eigenvalue-of-B}(a), we show that the numerical evaluation of \eqref{eq:beta_TAP} reproduces the prediction \eqref{eq:MF_Log_Temperature} from global mean-field theory, namely the logarithmic scaling of the transition temperature. In contrast to the global mean-field theory, however, the full connectivity structure is taken into account in the TAP approach.
Furthermore, we can see that the linearization \eqref{eq:TAP_linear} is consistent with the solution of the full TAP equations \eqref{eq:TAP}. Indeed, the projection of the magnetization $\mathbf{m}$ onto the leading eigenvector $\mathbf{v}_{\mathbf{B}(\beta),\mathrm{max}}$ of $\mathbf{B}(\beta)$, 
\begin{equation}
p_{\mathbf{B}(\beta)}=\frac{\mathbf{v}_{\mathbf{B}(\beta),\mathrm{max}} \cdot \mathbf{m} }{|\mathbf{m}|}\, \theta(|\mathbf{m}|)\label{eq:ProjB}
\end{equation} at the transition temperature  approaches unity, as seen in \prettyref{fig:Leading-eigenvalue-of-B}(b).

We will now explain the logarithmic scaling of the transition temperature in local TAP theory by examining the role of the additional diagonal term $-\beta\delta_{ij}k_i$ in \eqref{eq:def_B}. For given values of $m_{i}$, the presence of node $i$ affects the network like a heterogeneous
external field which couples most strongly to the nearest neighbors of $i$. We thus expand $m_{j}$ around $m_{i}=0$: 
\begin{equation}
m_{j}=m_{j}\big|_{m_{i}=0}+\beta A_{ji} m_{i} +\mathcal{O}(\beta^2) \label{eq:HubFields}
\end{equation}
keeping only terms up to linear order in $m_{i}$, $m_{j}$, and $\beta$. We then insert this expression into \eqref{eq:TAP_linear} to obtain
\begin{equation}
m_{i}=\beta\sum_{j}\bigg[A_{ij}\bigg(m_{j}\big|_{m_{i}=0} + \beta A_{ji} m_{i} \bigg) - \beta\delta_{ij}k_{i}m_{i}\bigg]\, ,\label{eq:Expanded_NN}
\end{equation}
up to quadratic order in $\beta$, consistent with the high-temperature expansion employed in \prettyref{eq:TAP}. Using $\sum_j A_{ji} = k_i$, this simplifies to
\begin{equation}
m_{i}=\beta\underbrace{\sum_{j}A_{ij}m_{j}\big|_{m_{i}=0}}_{\text{field in the absence of }i},\label{eq:LooseLocalInfo}
\end{equation}
which is similar to the linearized mean-field equation \prettyref{eq:MF_linear}, but is consistent with the notion of global order: the field at a node $i$ has no local contribution from the value of $m_i$, as it is calculated in the absence of $i$. This cancellation has been observed in previous studies \cite{fischer_spin_1991}, and can be anticipated from its relation to the fluctuation-dissipation theorem and cavity methods, which we discuss in \prettyref{app:Fluctuation-dissipation-theorem}.

\begin{figure}
\begin{centering}
\includegraphics{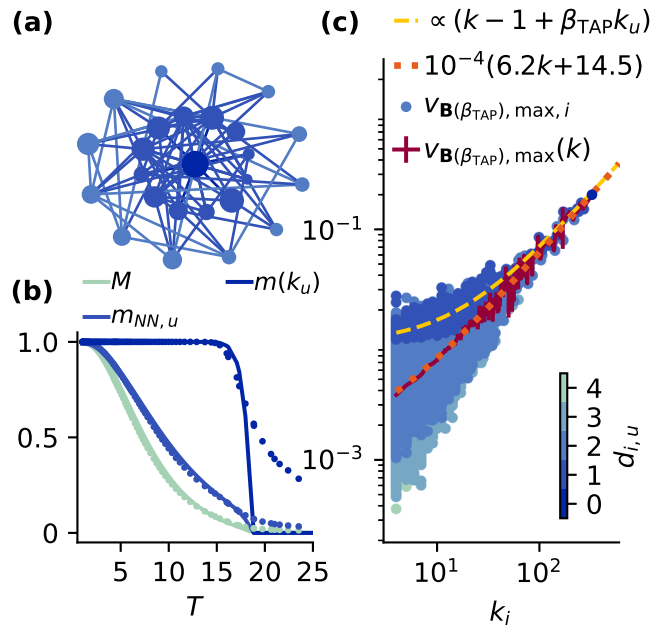}
\par\end{centering}
\caption{
(a) Hub $u$, the node with the largest degree, shown as central dark dot. Subset of the first $25$ nodes $i$ of the network, colored according to the minimal number $d_{i,u}$ of edges between $i$ and hub $u$. (b) Average  magnetization $M$ \eqref{eq:M(S)}, magnetization $m_{\mathrm{NN},u}$ \eqref{eq:M_NN(S)} averaged over the nearest neighbors of $u$, and magnetization $m(k_u)$ \eqref{eq:m(k)} of the hub as functions of the temperature $T$ within the TAP approach for a BAI model. Symbols denote the corresponding Monte Carlo data. (c) Eigenvector of the largest eigenvalue of $\mathbf{B}(\beta)$ at $T=\frac{m_{0}}{2}\,\log(N)$. Dots: entries of the eigenvector, sorted by their degree $k_i$, and colored according to  the distance $d_{i,u}$ to the central hub $u$. Red curve: Average of the entries with nodes of  equal degree. Dotted orange curve: linear fit of the averaged entry per degree. Dashed yellow curve: Scaling \eqref{eq:h_NN(S)(k)} of entry of a nearest neighbor of the
central hub $u$. The common prefactor is calculated as $\beta_{\mathrm{TAP}} S = k_u^{-1} v_{\mathbf{B}(\beta_{\mathrm{TAP}}),u}$. All three panels show data from the same BAI model with $N=10^4$, $m_0=4$.\label{fig:Spatial-heterogeneity}}
\end{figure}

Equation~\prettyref{eq:LooseLocalInfo} implies that the resulting eigenvector of $\mathbf{B}(\beta)$ cannot be localized strongly at one node of high degree, because the corresponding component cannot be influenced by its own presence. 
The TAP term $-\beta^{2}k_{i}m_{i}$ in \prettyref{eq:Expanded_NN} cancels the positive feedback of node  $i$ back to itself via any of its $k_i$ nearest neighbors $j$. As a result, \prettyref{eq:LooseLocalInfo} justifies the exclusion of strongly localized solutions by the replacement of $A_{ij}$ by $p_c(k_i, k_j )$, which is  used in the global mean-field theory. By analogous calculations, one then obtains the same transition temperature as in global mean-field theory, i.e., 
\begin{equation}
T_{\mathrm{TAP}}= \frac{\langle k^{2}\rangle}{\langle k\rangle}\approx\frac{m_{0}}{2}\log(N),\label{eq:TransitionTemperature}
\end{equation}
where the magnetization of each node depends on its degree as
\begin{equation}
m(k)=kS. \label{eq:TAP_LinearDegreeScaling}
\end{equation}
The description of the transition behavior in terms of the degree resolved quantities in this global TAP theory is therefore identical to the global mean-field theory. Numerically, we observe that the transition temperature obtained from this global TAP approach is consistent with the emergence of the non-trivial solution on the local level, namely the temperature at which \prettyref{eq:beta_TAP} is fulfilled, as shown in \prettyref{fig:Leading-eigenvalue-of-B}(a). We show in \prettyref{fig:Spatial-heterogeneity}c that the entries in the leading eigenvector of $\mathbf{B}(\beta_{\mathrm{TAP}})$, when averaged over nodes of the same degree, agree with the expected linear behavior in \eqref{eq:TAP_LinearDegreeScaling}.

In summary, the TAP term in \prettyref{eq:TAP} resolves the inconsistency between the local and the global approach that is present in mean-field theory.
This allows us to describe the magnetization beyond the degree-resolved approximation. 
For the nearest neighbors of a node $i$, \eqref{eq:HubFields} explicitly shows the influence of node $i$ on its neighbors, which we call the local bias. Thus, the local information has not been lost completely: the presence of node $i$ is encoded in the fields of its nearest neighbors. This is the reason why nodes of low degree show strong variations around the prediction \eqref{eq:TAP_LinearDegreeScaling}. Their magnetization reflects a more complicated local structure. This is seen explicitly when considering  the nodes according to their respective distance $d_{l,u}$ to a hub $u$, as shown in \prettyref{fig:Spatial-heterogeneity}(a) for the example of the hub $u$  of largest degree. The local bias due to the strong alignment of the hub with the overall magnetization  affects its nearest, next-nearest, and further neighbors, forming thereby an onion-like structure, where in each shell the bias from the hub becomes weaker. This structure is also visible in the eigenvector of $\mathbf{B}(\beta)$ at $\beta = \beta_{\mathrm{TAP}}$, as shown in \prettyref{fig:Spatial-heterogeneity}(c): Nodes of low degree that are not in the vicinity of any hub are much more weakly magnetized than those that are nearest neighbors to a hub.
Nodes of high degree connect to a representative collection of nodes on the network and therefore do not deviate much from the expectation value $m(k)$, as seen in \prettyref{fig:Spatial-heterogeneity}c.  
As a consequence of the hierarchical network structure, we observe that the transition marks the point below which hubs essentially become permanently ``frozen'', i.e. they almost perfectly align with the overall magnetization of the whole system, $m_{u}\approx1$ for $k_{u}\gg\langle k\rangle$ (see \prettyref{fig:Spatial-heterogeneity}(b)).  

The cancellation of the self-feedback \prettyref{eq:LooseLocalInfo} also extends to lower temperatures. Using the same expansion method, one finds that up to quadratic terms in the inverse temperature $\beta$, the self-consistency equation reads
\begin{equation}
m_{i}\approx\tanh \Big(\beta\,\sum_{j}A_{ij}m_{j}\big|_{m_{i}=0} \Big). \label{eq:Cancelled_Selffeedback_below_T_TAP}
\end{equation}
To understand the degree-resolved magnetization, we insert $m_{i}\approx m(k_{i})$ into  \eqref{eq:Cancelled_Selffeedback_below_T_TAP}, making use of the global theory also below the transition temperature. This leads to the approximate result for the degree-resolved average magnetization from the global mean-field theory~\citep{bianconi_mean_2002,dorogovtsev_ising_2002,leone_ferromagnetic_2002},
\begin{equation}
  m(k) \approx \tanh \left(\beta\,k\,S \right). \label{eq:m(k)}
\end{equation}
This approximation may not be good for small $k_{i}$, but nodes of larger degree contribute more strongly to the global order parameter $S$, for which $m_{i}\approx m(k_{i})$ holds quite precisely.
The resulting  equation for the global order parameter
\begin{align}
S & =\frac{1}{\langle k\rangle}\sum_{k} p(k)\,k\,\tanh\big(\beta\,k\,S\big)
\label{eq:Defintion_S}
\end{align}
can be solved numerically for arbitrary network sizes. From this, we obtain the total magnetization
\begin{equation}
  M = \sum_{k}p(k)\tanh\left(\beta\,k\,S\right), \label{eq:M(S)}
\end{equation}
which compares well to the Monte Carlo data (\prettyref{fig:Spatial-heterogeneity}(b)). 
At the transition, obtained from either TAP or global mean-field theory, we find that the prediction $m_{i}\propto k_{i}$ averaged over all vertices of the same degree holds for small degrees (see \prettyref{fig:Strong-local-differences-in-magnetization}(a)), albeit at a slightly downward shifted effective transition temperature (cf. \prettyref{sec:Finite-size-effects} for a discussion on the exact position of this crossover).
For large degrees, however, the average behavior in the Monte Carlo data seems to deviate from this rule. This can be understood from what is essentially a combinatorial argument. 
For finite systems, the modulus of the magnetization is never exactly zero, even in the disordered case, but rather converges to a finite value. This finite value only converges towards zero with increasing number of spins over which the average is performed. Even though we somewhat alleviate this effect by measuring with respect to the net magnetization, the finite size of the network still affects the ordering of the nodes of larger degrees (see \prettyref{fig:Strong-local-differences-in-magnetization}(b)):
firstly, they tend to align to the net magnetization, due to the large energetic cost involved in  flipping their orientation, so that in the limit of large $k_{i}$, equation \prettyref{eq:MC_mag_h} effectively reduces to the modulus again. Secondly, by virtue of having a high degree, these nodes tend to be few. Keeping these two points in mind, one can understand the deviation from the $m_{i}\propto k_{i}$ scaling for large $k_i$ by this finite magnetization converging to a plateau as the magnetization eventually has to saturate to unity. One can also understand the degree-dependent finite values of the local magnetization in the limit of high temperatures as seen in \prettyref{fig:Spatial-heterogeneity}.
The local structure however, remains visible across all temperature scales. Most prominently, the magnetization at the nearest neighbors of the hubs is enhanced as compared to the average value in \prettyref{fig:Spatial-heterogeneity}(b).
Generalizing equation \prettyref{eq:HubFields}, we can also calculate the average magnetization of nearest neighbors of a hub $u$. The effective field is the sum over all neighbors, therefore for a node of degree $k$, we find one contribution from $m(k_u)$, and the remaining $k-1$ connections couple to $S$, so the effective field reads
\begin{equation}
h_{\text{NN},u}(k) = \beta \,(k-1)\,S +\beta \, m(k_{u}) \,. \label{eq:h_NN(S)(k)}
\end{equation}
In the linear regime, where $m(k_u) =\beta_{\mathrm{TAP}} k_u S $ and $m_{\text{NN},u}(k)= h_{\text{NN},u}(k)$ we find that this equation well describes the entries for the nearest neighbours of the hub $u$ in the leading eigenvector of $B(\beta_{\mathrm{TAP}})$ in \prettyref{fig:Spatial-heterogeneity}c.
Making use of $p_c$ and averaging over all $k_u$ neighbors of $u$  yields
\begin{equation}
m_{\text{NN},u}=\frac{1}{\langle k\rangle}\sum_{k}p(k)\,k\,\tanh\big[\beta \,(k-1)\,S +\beta \, m(k_{u})\big], \label{eq:M_NN(S)}
\end{equation}
which predicts an elevated magnetization at those nodes as compared to the average value (\prettyref{fig:Leading-eigenvalue-of-B}(b)). We thus find that the global approach, used here within the TAP approach, can be extended to account for the local differences in the ordered state of the system in a consistent manner. 

\section{Effective magnetic transition on finite networks\label{sec:Finite-size-effects}}

\begin{figure}
\begin{centering}
\includegraphics{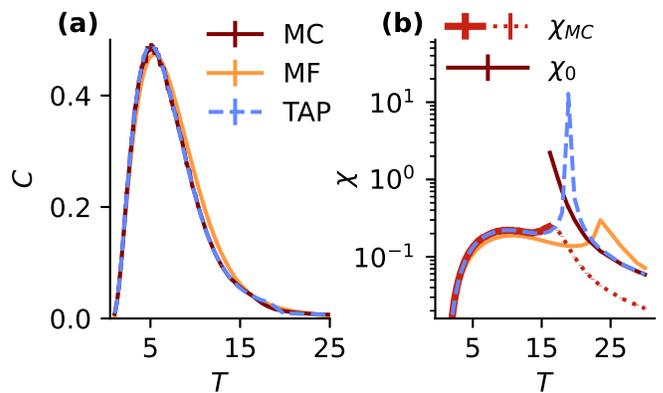}
\par\end{centering}

\caption{Specific heat $C$ (a) and 
 susceptibility $\chi$ (b)  as functions of temperature $T$ for a BAI with $N=10^{4}$ and $m_{0}=4$. $\chi_{\mathrm{MC}}$ \prettyref{eq:Susceptibility_MC} is shown as a full line below (overlaid by TAP) and as a dotted line above the Monte Carlo transition temperature. Monte Carlo data for $\chi_0$ \prettyref{eq:Susceptibility} above the effective transition temperature, which is the appropriate estimator for the susceptibility in  the paramagnetic regime. All 
data obtained for the same realization of the adjacency matrix $\mathbf{A}$ of the network, where $N=10^4$ and $m_0=4$.
\label{fig:Susceptibility-and-specific-heat}}
\end{figure}

We further probe the TAP approach by  calculating  the specific heat $C$, 
and the susceptibility $\chi$.
Here, the susceptibility measures the response to a global, homogeneous external field $h_i = h$ for all $i=1,...,N$. In TAP and mean-field theory, the susceptibility is obtained as the inverse of the Hessian of $G$ (see \prettyref{app:Plefka-expansion}), and the specific heat is computed from the numerical derivative of $E$. Both quantities are shown in \prettyref{fig:Susceptibility-and-specific-heat}. 

Both in mean-field and in TAP approximation the susceptibility exhibits a pronounced peak at the point where the non-trivial solutions to \prettyref{eq:MF} or \prettyref{eq:TAP}, respectively, come into existence  (\prettyref{fig:Susceptibility-and-specific-heat}(b)). This is readily understood from considering the manifold of possible solutions to the equation of state \eqref{eq:EquationOfState}: Above their respective transition temperature, both the mean-field and TAP equations allow only for a single trivial solution, for which the magnetization vanishes on all nodes, such that the expectation value of the absolute magnetization is exactly zero (see \prettyref{fig:M,E}). Below the transition temperature, the equations feature a pair of solutions that are transformed into one another
by the $Z_{2}$-symmetry. This is also the reason why within this approximation, the susceptibility diverges even for finite system sizes (see \prettyref{fig:Susceptibility-and-specific-heat}) right at the temperature for which   the  transition from a unique solution to a triplet of solutions takes place:
At the transition, $G$ must have a saddle point, therefore the inverse of its Hessian, i.e., the susceptibility, diverges.

Such a sharp peak is however absent in the Monte Carlo data,
as befits the exact solution on a finite system. 
Indeed, there cannot be any non-analytical behavior in the free energy on  a finite system at finite temperatures, and thus such sharp features cannot exist in the physical observables. Nevertheless, there appears a broad  peak in the estimator \eqref{eq:Susceptibility_MC} of the susceptibility for the symmetry broken regime. In all cases, we take the position of the  peak in the susceptibility as an estimator for the effective transition temperature --  this is the point at which the system is most sensitive to small perturbations. 

\begin{figure}
\begin{centering}
\includegraphics{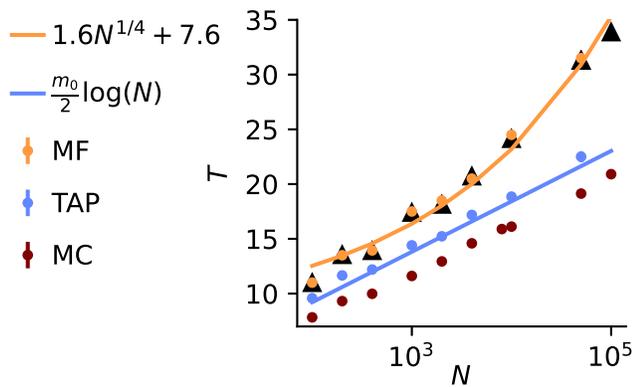}
\par\end{centering}
\caption{Effective transition temperature as extracted
from the maximum of the susceptibility as a function of system size
$N$ for $m_{0}=4$. Red dots: Estimated from Monte Carlo simulations using \prettyref{eq:Susceptibility_MC}. Yellow dots:
Solutions of the local mean-field equation \prettyref{eq:MF}. Violet dots:
Solutions of the TAP equation \prettyref{eq:TAP}.
Violet curve: Global mean-field  \eqref{eq:MF_Log_Temperature} or,  equivalently, TAP prediction \eqref{eq:beta_TAP}. Black triangles: Leading eigenvalue of the adjacency matrix $\lambda_{\mathbf{A},\mathrm{max}}$. Orange curve: Fit of $\lambda_{\mathbf{A},\mathrm{max}}$ to $a\,N^{{1}/{4}}+b$ to illustrate the $N^{{1}/{4}}$ scaling. \label{fig:Transition-temperature}}
\end{figure}

To further compare the susceptibility from Monte Carlo to the TAP solution, we must account for  the distinction between the symmetric and the symmetry-broken regime that we already mentioned  in \prettyref{sec:Monte-Carlo}. In the latter case, the second term in \prettyref{eq:Susceptibility} provides a
non-vanishing contribution, whereas in the former case, i.e., above the transition temperature, the magnetic susceptibility reduces to the
second moment of the magnetization, because $M=0$ in the non-symmetry broken regime. 
Based on  this distinction, we indeed find good agreement between the TAP and the MC susceptibility at both small and large temperatures, as can be seen in \prettyref{fig:Susceptibility-and-specific-heat}. While the agreement in the high-temperature regime is expected, as the TAP equations are an asymptotic expansion in $\beta$, the good agreement in the low-temperature regime is pleasantly surprising.
Only in the vicinity of the effective transition temperature do we observe notable differences between the TAP and Monte Carlo results. In fact, this concerns the value of the effective transition temperature $T_T$ itself, as TAP and Monte Carlo differ by a constant offset regarding the position of the maximum of the susceptibility. 
While this offset is still  visible up to large networks with  $10^5$ nodes and does apparently not reduce in absolute magnitude (cf. \prettyref{fig:Transition-temperature}), it nevertheless becomes negligible for large network sizes $N$, since in any case  the value of $T_T$ diverges logarithmically with $N$. 
In \prettyref{fig:Transition-temperature}, we show that in the mean-field and TAP approximation, we find that the transition temperature coincides with the expected result from the linear analysis, \eqref{eq:MF_True_T_T} and \eqref{eq:beta_TAP}, respectively. Monte Carlo simulations confirm the logarithmic scaling \prettyref{eq:MF_Log_Temperature}
obtained from global mean-field theory or from TAP theory \prettyref{eq:TransitionTemperature} up to a constant shift. This is consistent with previous work \cite{herrero_ising_2004,aleksiejuk_ferromagnetic_2002}.

The specific heat $C$ shows a distinct behavior from the susceptibility, with a maximum that is not related to the position of the effective transition temperature. Indeed, an analysis of its position for different system sizes confirms that the position of the maximum in $C$  as well as the value of the maximum is size-independent and already explained reasonably well within mean-field theory. In particular, we find the position of the maximum  to linearly shift to larger temperatures upon increasing the network parameter $m_0$.  This  can be anticipated  already within mean-field theory: Solving the linearized self-consistency equation in the continuum limit one obtains $T_{max, C} = 2 m_0$, which exhibits no $N$-dependence (even though the predicted value is quantitatively off). Overall, we observe a good agreement between the Monte Carlo data for the specific heat and the TAP results.
We note that \citep{dorogovtsev_ising_2002} arrives at a similar expression for the specific heat. Physically, the maximum in the specific heat $C$ shows that a substantial amount of entropy is released upon the full ferromagnetic alignment of the spin from a large number of nodes well below the initial onset of the ferromagnetic order, which, as discussed above, is an effect that is driven by the much lower number of highly-connected hubs in the system.

\section{Discussion\label{sec:Discussion}}

%
We investigated the onset of magnetic order in a system of Ising spins that interact on an undirected Barabási-Albert network. These networks are characterized by heavy-tailed degree distributions and thus possess hubs, nodes with a large degree of connectivity to other sites of the network.
The transition from a disordered to an ordered state can be understood within mean-field theory of a global order parameter \citep{bianconi_mean_2002}. Monte Carlo simulations confirm that an effective magnetic transition occurs at an effective transition temperature that scales with $\log(N)$, in line with mean-field theory.
The global mean-field theory, however, neglects the local structure of the network that extends beyond the distribution of the nodes' degree. This results in two shortcomings:
First, nodes with the same degree are predicted to have the same magnetization; Monte Carlo simulations, however, show this assumption to be violated, in particular for the neighboring nodes of the hubs.
Second, we find that the  mean-field theory for the global order parameter is inconsistent with the underlying mean-field theory for the magnetization of individual spins. This inconsistency is disturbing, because the former theory is derived from the latter.
Here, we resolved these shortcomings upon including the leading systematic perturbative correction in the interaction strength to the mean-field approximation. This correction cancels the spurious, indirect self-feedback from one node onto itself, mediated by its direct neighbors, which is still present in mean-field theory. We identify this self-feedback as the cause of the inconsistency between the two mean-field approaches; the corrected mean-field (TAP) approach is indeed consistent between its local and its global formulation.

This improved theoretical understanding sheds light into the mechanism driving
the transition into the heterogeneous system state. In particular, the theory qualitatively reshapes the role that hubs and their direct neighbors play for the onset of order in two ways: 
Hubs enforce a stronger alignment of their nearest neighbors as compared to the mean-field prediction, which accounts for only their neighbours' degrees.
At the same time, local mean-field theory overestimates the importance of hubs for the build-up of order in the first place.
The reason is a spurious self-feedback; if it was present, it would endow individual hubs to self-stabilize their magnetization and thus drive the transition already at temperatures 
$T \propto N^\frac{1}{4} \gg \log(N)$. The cancellation of the self-feedback by the perturbative corrections thus explains why hubs are less effective in driving global order than may be naively expected.
Likewise, nodes in  direct vicinity of hubs show stronger local order than predicted by their degree alone. The reason for this effect are the local fields that are stronger than expected by global mean-field theory. 

The TAP equations take self-feedback - mediated by nearest neighbors - into account. However, such feedback can also come from loops in the connectivity structure, due to the  overnext and over-overnext nearest neighbours and so on in a systematic high-temperature expansion \citep{georges_how_1991, kuhn_expansion_2018, vasiliev_legendre_1974}. 
In principle, an infinite number of corrections can be taken into account. It may be worthwhile to explore even higher order corrections to find if this self-consistent theory may explain the constant downward shift between the Monte Carlo and TAP results for the effective transition temperature. However, we expect these corrections due to higher order terms to be small, given  (i) the small prefactor $\beta$ and (ii) the good overall agreement between TAP theory and the  Monte Carlo simulations. 

The finite size of the system is taken into account by solving the TAP equations with respect to the full adjacency matrix. It should be noted that in the limit of large $N$, the transition temperature shifts to
infinite temperatures, so that the system is always in an ordered state in the thermodynamic limit. This defies the notion of regular phase transitions that would normally only acquire meaning in the thermodynamic limit.

To our knowledge, this work constitutes a successful effort to also account for the  local structure in finite size networks, while global quantities such as the transition temperature, and the influence of network parameters,
such as the clustering coefficient and average connectivity, have been investigated quite thoroughly in the past \citep{leone_ferromagnetic_2002,dorogovtsev_ising_2002,herrero_ising_2015,bianconi_mean_2002}. In comparison to \citep{dorogovtsev_ising_2002,leone_ferromagnetic_2002,bianconi_mean_2002}, our method allows us to  compute  averages using the full connectivity structure, whereby local effects such as the bias fields become visible.
The cancellation of self-feedback by perturbative corrections is reminiscent of the cavity method \citep{mezard_spin_1986}; here, too, the field sensed by a spin needs to be computed in the absence of this very spin. The cavity method is prominent in the analysis of spin glasses.

Irrespective of the particular network architecture, we expect the cancellation of self-feedback to hold
in Ising models, enabling one to find the transition temperature by means of a linear problem. We expect the  distinction between local structure (for example, the immediate neighborhood of a specific
node) and global hierarchy to be beneficial to the analysis of similar
problems. In particular, we expect that qualitatively similar mechanisms on the local and on the global scale as described here shape the emergence of order in general heterogeneous systems.

\section{Acknowledgments}
We thank A. Fischer and J. Krishnan for helpful discussions. We acknowledge the IT Center at RWTH Aachen University and the JSC Jülich for access to computing time through JARA-HPC, and acknowledge support by the DFG through RTG 1995, by RWTH Exploratory Research Space Seed Funds, by the JARA Center for Doctoral studies within the graduate School for Simulation and Data Science (SSD), and by the Excellence Strategy of the Federal Government and the Länder (G:(DE-82)EXS-PF-JARA-SDS005).

\bibliographystyle{apsrev4-2}
\bibliography{bibliographySmall}

\onecolumngrid

\appendix

\section{Construction algorithm \label{app:Construction-algorithm}}

We briefly state the construction algorithm of a BA network  with $N$ nodes,
which depends on a parameter $m_{0}$: Starting with $m_{0}\in\mathbb{N}$
unconnected nodes with labels $1,\dots\,m_{0}$, we connect to
each of those nodes one additional node, labeled $m_{0}+1$. The remaining
$N-m_{0}-1$ nodes are added iteratively to the already existing network, each connecting
to $m_{0}$ randomly chosen existing nodes, where the probability
to connect a new node $j$ to a node $i<j$ is proportional to
the degree $k_{i}:=\sum_{j}A_{ij}\geq m_{0}$ (i.e, the number of connections to node 
$i$), so that $p_{ij}\propto k_{i}.$ In the large-$N$ limit, to a good approximation,
the total number of connections on the network is given by $Nm_{0}$, so that the average degree is 
\begin{equation}
\langle k\rangle=\frac{2m_{0}N}{N}=2m_{0},
\end{equation}
because each connection joins two nodes. 

\section{Plefka expansion\label{app:Plefka-expansion}}

Here, we state how to obtain the TAP equations \prettyref{eq:TAP}
by means of the Plefka expansion \citep{plefka_convergence_1982}.
The name TAP theory originates from \citep{thouless_solution_1977},
where the expressions are presented as a fait accompli. The first derivation has been given in \citep{vasiliev_legendre_1974}. Alternative
ways to arrive at this result, including diagrammatic expansion methods,
may be found in \citep{georges_how_1991,kuhn_expansion_2018}.

The strategy is to expand \eqref{eq:G} around the non-interacting case, where the coupling term in the energy vanishes and $G$
can be computed exactly. The advantage of expanding $G$ and not $F$ is that $G$ describes the distribution for given mean values $m_i$ and thus yields implicit expressions for their values
using \prettyref{eq:EquationOfState}. Furthermore, the diagrammatic
expansions of $G$ contains fewer terms (see \citep{vasiliev_analysis_1975,kuhn_expansion_2018}) than the expansion of $F$.
We now split the Hamiltonian into an interacting part $H_{\text{int}}$ and a non-interacting part, re-introducing the interaction
strength $J$,

\begin{equation}
H=\underbrace{- \frac{J}{2}\sum_{ij}A_{ij}x_{i}x_{j}}_{=:J H_{\text{int}}}-\sum_{i}h_{i}x_{i}\,
\end{equation}
and proceed by expanding $G$ in $J$ to second order.  Here, for $J=0$ , i.e., the non-interaction case, is simply the $G_{J=0}$
entropy of independent binary variables with mean values $m_{i}$,

\begin{equation}
\beta G_{J=0} = \sum_{i} \frac{1+m_{i}}{2}\log\left(\frac{1+m_{i}}{2}\right)+\frac{1-m_{i}}{2}\log\left(\frac{1-m_{i}}{2}\right)\:.
\end{equation}
The first derivative provides the expectation value of the interaction energy, which, when evaluated for independent variables, and fixed values of 
$m_{i}$, gives
\begin{align}
\partial_{J} \beta G\bigg|_{J=0}=& \beta \big\langle H_\mathrm{int} \big\rangle\\
 = & -\frac{\beta}{2} \sum_{ij}A_{ij}\langle x_{i}x_{j}\rangle_{J=0}=-\frac{\beta}{2}\sum_{ij}A_{ij}m_{i}m_{j}\,.
\end{align}

To obtain  the second derivative, one has to evaluate
\begin{equation}
\partial_{J}^{2} \beta G\bigg|_{J=0}=\beta^2 \Bigg\langle H_{\text{int}}\bigg[H_{\text{int}}-\big\langle H_{\text{int}}\big\rangle-\sum_{i}(x_{i}-m_{i})\partial_{J}h_{i}\bigg]\Bigg\rangle\bigg|_{J=0}\,,
\end{equation}
where the derivative of $\partial_{J}h_{i}$ can be found using the inverse of
\eqref{eq:EquationOfState} and inserting $G \approx G\big|_{J=0}+ \partial_{J}G\big|_{J=0} J $. Summing all contributions and finally setting $J=1$ again, one obtains \eqref{eq:G_expanded} in the main text.
For the susceptibility, we make use of the fact that $\nabla_{h}F$
and $\nabla_{m}G$ are (up to a sign) inverse functions of each other:
\begin{equation}
\partial_{h_{i}}\left(\nabla_{m}G\bigg|_{m_{i}=-\partial_{h_{i}}F}\right)_{j}=\sum_{k}\frac{d^{2}G}{dm_{j}dm_{k}}\frac{d^{2}F}{dh_{i}dh_{k}}=-\delta_{ik}\,.
\end{equation}
After having determined $m_{i}$ for all $i$, using \prettyref{eq:TAP}, we can
therefore obtain the susceptibility $\chi_{ij}=\frac{dm_{i}}{dh_{j}}=-\frac{d^{2}F}{dh_{j}dh_{k}}$
by matrix inversion of the Hessian of $G$. 
However, the average response to a global field (i.e., $h_i = h$ for all $i$)  is given by  the sum over all entries
\begin{equation}
    \chi = N^{-1} \sum_{ij} \chi_{ij}.
\end{equation}
One can verify by insertion that in order to determine the average susceptibility, one needs to  simply solve 
\begin{equation}
G^{(2)}\boldsymbol{\overline{\chi}}=N^{-1}\left(1,1,\dots,1\right)^{\T}\,
\end{equation}
for $\boldsymbol{\overline{\chi}}$ and finally perform the summation  over all entries,
\begin{equation}
    \chi =\sum_{i} \overline{\chi}_i.
\end{equation}

\section{Self-averaging on BA networks \label{app:Self-averaging-of-BA-networks}}

\begin{figure}
\includegraphics{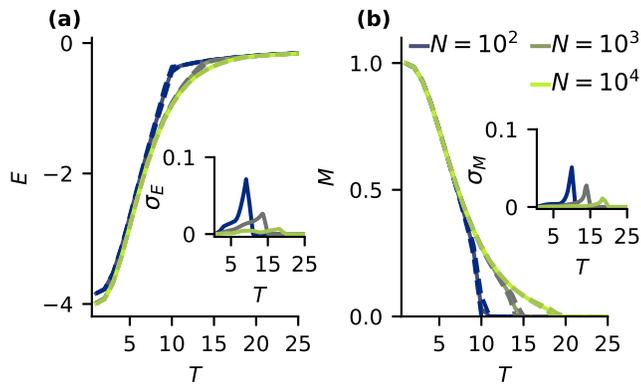}

\caption{ Average energy $E$ (a) and magnetization $M$ (b) from ten different realizations of the BAI model, as functions of temperature $T$ for different system sizes $N$ and $m_{0}=4$ obtained from TAP approximation. Dashed lines are drawn one standard deviation above and below the mean, calculated according to \prettyref{eq:Standard_Deviation_Realizations}. Lines seem to align where the standard deviations are very small. Insets show the standard deviations alone.\label{fig:Self-averaging}}
\end{figure}

Here we show that global properties, such as the  magnetization $M$ and the energy $E$ do
not depend significantly on the specific  realization of the BA networks. To this end,
we calculate within the TAP approach the variances  $\sigma_{O}$ of such an  observable $O$, with

\begin{equation}
\sigma_{O}=\sqrt{\langle O^{2}\rangle_{R}-\langle O\rangle_{R}^{2}},\label{eq:Standard_Deviation_Realizations}
\end{equation}
where $\langle\cdot\rangle_{R}$ denotes the average over independent realizations
of the BA network with given values of $N$ and $m_0$. The results are shown in \prettyref{fig:Self-averaging}. With
increasing system size, neither the magnetization nor the energy show a significant variance, and furthermore, the variances decrease with increasing system size. 

\section{Details on the Monte Carlo simulation scheme \label{app:Monte-Carlo}}

When treating the  Ising model on a network with hierarchical
structure, such as a BA network, several considerations are required in order to choose
an appropriate simulation method.
When performing ordinary Metropolis Monte Carlo simulations, one quickly
runs into the problem of ``freezing hubs''. The local update step
is increasingly unlikely to flip nodes of high connectivity, due to
the massive change of energy that is involved in such a single spin flip. The problem
is akin to the emergence of domain walls in a regular Ising model
on a lattice, although it quickly gets more severe as the degree of
the hubs grows with increasing system size. With the hub frozen out,
 Metropolis updates are no longer able to sample the complete phase
space, autocorrelations exhibit long time scales, and the measurement of physical
observables is prone to large statistical errors or the effective breakdown of ergodicity. 

To circumvent this problem in the dynamics of the simulation, the results
obtained by Monte Carlo sampling reported here use a slightly modified version of
the parallel-tempering method on the basis of the work of \citep{swendsen_replica_1986},
as well as feedback-optimized methods first developed by \citep{katzgraber_feedback-optimized_2006},
in combination with regular Metropolis update steps. After initializing
a given BA network with a random initial distribution of the Ising
spins, one creates $M$ replicas of the given grid and evolves them
separately at different temperatures, according to a local stochastic
process, i.e. the Metropolis-Hastings algorithm. Then, periodically after a fixed
amount of these local update steps, one checks for valid swaps between
neighbors of replicas in temperature space, according to an appropriate pairwise
transition probability $p_{ij}$, which is determined by the state of
the two replicas. If such an update step is accepted, the algorithm
swaps the spin realizations of the two replicas, thus effectively allowing
each replica to move through the full temperature space. Initially designed for
the simulation of spin glasses, this sampling method allows us to overcome large energy barriers that would normally make the sampling difficult. More specifically, in the case of a BA network,
the most prominent of these energy barriers is of course the flip
of a ``hub''. Employing parallel tempering allows us to reduce autocorrelation
times and recover an ergodic sampling  within each single Monte Carlo simulation. A
global update step constructed in this way, i.e. the swapping of two
replicas between different temperatures, must of course fulfill the
detailed balance condition of the overall Markov chain, and thus the swap probabilities
are chosen to be
\begin{align*}
p_{ij}= & \min(1,e^{(E_{i}-E_{j})(\beta_{i}-\beta_{j})}).
\end{align*}

This of course raises the question on how to distribute the $M$ different temperatures
in $[T_{0},T_{M}]$ used in a parallel-tempering simulation. There exists
a multitude of schemes to tackle this problem in the literature, each
pursuing different objectives with regards to the simulation dynamics.
The interpretation of replicas moving through temperature space allows
for an adequate physical picture in terms of the replicas' diffusivity along the  temperature
axis. In order to optimize the gain in reduced autocorrelations  from
the parallel tempering method, one thus tries to maximize the number of
full ``round-trips'' that a replica undertakes. In order to make this
feasible, one tags each replica moving in temperature space as ``up''
when it reaches the minimum temperature and with ``down'' when it
reaches the maximum temperature. One can then increment a temperature-wise
histogram after every parallel tempering update step, by counting the number
of up-walking and down-walking replicas at a given temperature. We
keep track of this distribution by defining
\[
f(T_{i})=\frac{n_{\mathrm{up}}}{n_{\mathrm{down}}+n_{\mathrm{up}}}(T_{i}).
\]

In \citep{katzgraber_feedback-optimized_2006} it is shown,
that, in order to guarantee a maximum number of round-trips, this quantity
is supposed to be a linearly descending series. Since $f$ is monotonous,
we can define an inverse $g$ of $f$, such that
\[
g(f(T_{i}))=T_{i}
\]
Feeding a linearly descending series to this $f$ will create a new set
of temperatures, which is then used in the next preliminary parallel tempering
simulation, until the temperature grid converge and an optimal distribution
of the simulation temperatures is reached. However, the originally proposed method
ran into problems when applied to large BA  networks. The
shifting of temperatures in the self-optimization proved to be too
severe, which results in the algorithm getting trapped between ever
newly created diffusivity bottlenecks. As an extension of the original
method, we thus define
\begin{align*}
f_{\omega}(T_{i})= & (1-\omega)f(T_{i})+\omega(1-\frac{i}{M}),
\end{align*}
where $\omega\in[0,1]$. This addition effectively smooths out the
rearranging of temperatures and guarantees a converging optimization.
Usually $\omega=0.75$ was chosen and a few temperatures were manually
assigned after the optimization in order to obtain a better resolution around physically
interesting points. At the end of the procedure, we obtain a Monte Carlo
algorithm that effectively samples the full phase space and produces
more precise data using less computational resources.

\section{Fluctuation-dissipation theorem\label{app:Fluctuation-dissipation-theorem}}

In mean-field theory, the presence of the spin at node $i$ may be alternatively viewed as an inhomogeneous external field
of strength $h_{\mathrm{ext},j}=A_{ji}m_{i}$ at node $j$.
The response to this field on the spin at node $j$ is given by $\chi_{j}A_{ji}m_{i}$, where
$\chi_{j}$ is the the susceptibility at node $j$, given by

\begin{equation}
\chi_{j}=\frac{\partial m_{j}}{\partial h_{\mathrm{ext},j}}=\beta\,(1-m_{j}^{2}).\label{eq:Susceptibility_appendix}
\end{equation}
Up to factors of $\beta$, this is equal to the variance

\begin{equation}
\langle x_{j}^{2}\rangle-\langle x_{j}\rangle^{2}=1-m_{j}^{2},\label{eq:Variance}
\end{equation}
in accord with  the fluctuation-dissipation theorem \citep{goldenfeld_lectures_1992}. When sampling from an equilibrium distribution, the fluctuation-dissipation theorem applies, which states the equivalence
between the fluctuation \prettyref{eq:Variance} and the linear response
\prettyref{eq:Susceptibility_appendix}. In fact, both values are calculated
in the same way, namely (up to factors of $\beta$) from the second derivative
of the free energy $F$ with respect to the external field $h$.

The self-feedback of the response field $\chi_{j}A_{ji}m_{i}$ at node 
$j$ to the  moment at node $i$ is given by

\begin{equation}
A_{ij}^{2}\beta^{2}m_{i}(1-m_{j}^{2}).\label{eq:LinearResponse}
\end{equation}
Summing over all neighbor nodes $j$, this gives precisely the TAP
term in \eqref{eq:TAP}, but with opposite sign.

For non-equilibrium systems (the non-equilibrium kinetic Ising model
or directed networks of binary units), this equivalence between linear response and fluctuations is lost \cite{roudi_dynamical_2011},
as this argument does not produce the correct time arguments of the
mean fields and the replacement $A_{ij}=A_{ji}$ is no longer possible. This was also noted in \citep{kappen_mean_2000}. Therefore, the cancellation of the spurious self-feedback present in the mean-field approximation by the TAP correction term cannot be expected outside thermal equilibrium.

\end{document}